\documentclass[abbrv,aps,prl,twocolumn,showpacs,preprintnumbers,amsmath,amssymb,
superscriptaddress]{revtex4}

\usepackage{amsmath,amssymb,amsfonts}
\usepackage{bm,graphicx,color}

\newcommand{\laco}[1]{$\mathrm{LaCoO_3 }$}

\begin{document}
\title{Spin state of negative charge-transfer material SrCoO$_3$}
\author{J. Kune\v{s}}
\affiliation{Institute of Physics, Academy of Sciences of the Czech Republic, Cukrovarnick\'a 10,
Praha 6, 162 53, Czech Republic}
\author{V. K\v{r}\'apek}
\affiliation{Institute of Physics, Academy of Sciences of the Czech Republic, Cukrovarnick\'a 10,
Praha 6, 162 53, Czech Republic}
\author{N. Parragh}
\affiliation{Institut f\"ur Theoretische Physik und Astrophysik, Universit\"ut W\"urzburg, 
Am Hubland, D-97074 W\"urzburg, Germany}
\author{G. Sangiovanni}
\affiliation{Institut f\"ur Theoretische Physik und Astrophysik, Universit\"ut W\"urzburg,
Am Hubland, D-97074 W\"urzburg, Germany}
\author{A. Toschi}
\affiliation{Institut f\"ur Festk\"orperphysik, Technische Universit\"at Wien, Vienna, Austria}
\author{A.~V. Kozhevnikov}
\affiliation{Institute for Theoretical Physics, ETH Zurich, CH-8093 Zurich, Switzerland}
\date{\today}

\begin{abstract}
We employ the combination of the density functional and the dynamical mean-field theory (LDA+DMFT)
to investigate the electronic structure and magnetic properties of SrCoO$_3$,
monocrystal of which were prepared recently. 
Our calculations lead to a ferromagnetic metal in agreement with experiment. We find
that, contrary to some suggestions, the local moment in SrCoO$_3$ does not arise
from intermediate spin state, but is a result of coherent superposition of many different
atomic states. We discuss how attribution of magnetic response to different 
atomic states in solids with local moments can be quantified.

\end{abstract}
\maketitle

The microscopic origin of the paramagnetic (PM) moment is one of the key questions
in materials with strongly correlated electrons. In particular, local moments with
magnitudes far from any atomic limit pose a non-trivial problem. 
The perovskite cobaltites Sr$_x$La$_{1-x}$CoO$_3$, an example of such system, 
attracted much attention. 
The variety of possible valence and spin states of the Co ion and their
nearly degenerate energies are behind
the strongly temperature ($T$) and pressure dependent magnetic susceptibility 
and conductivity in LaCoO$_3$ \cite{laco} as well as
the formation of large magnetic polarons and spin glass in Sr$_x$La$_{1-x}$CoO$_3$ at small doping $x$. 
At larger dopings the material becomes a ferromagnetic (FM) metal \cite{srla}
and remains so up to the stoichiometric SrCoO$_3$ composition \cite{long2011}.
Fractional PM moment in LaCoO$_3$ is traditionally thought to arise
from the statistical mixture of the low spin (LS) ground state and high spin (HS)
or intermediate spin (IS) excited state. While the IS state cannot
be the ground state of an isolated ion in a crystal-field, it was suggested that it
may be stabilized by the covalent Co-O bonding \cite{goodenough71,potze95,korotin}.
SrCoO$_3$ is considered a candidate for realization of IS ground state \cite{potze95,zhuang98}.

Is it possible to associate the PM moment with a particular atomic state
or states in materials with strong metal-ligand hybridization? Can we quantify
the contributions of different states? 
We address these questions in the case of SrCoO$_3$ and propose a general
way to answer them. To this end we employ the LDA+DMFT 
approach \cite{ldadmft} 
to compute the k-resolved spectral functions, the reduced density matrix for the Co site, 
and local two-particle correlation functions including the local spin susceptibility. 
SrCoO$_3$ is found to be metallic both in PM and FM
phase. We show that its PM moment cannot be 
associated with either IS or HS state of Co.

The calculation proceeds in several steps. First,
an LDA calculation for the experimental perovskite structure is performed
using WIEN2k \cite{wien2k} 
density functional code.  
The converged bandstructure is represented in the Wannier function
basis \cite{wannier} spanning the Co $d$ and O $p$ bands. 
The averaged screened Coulomb and exchange parameters for the Co-$d$ orbitals $U$=10.83~eV and $J$=0.76~eV  were
derived using the constrained RPA technique \cite{cRPA}. Following the
ideas of Ref.~\onlinecite{cRPApd}, we screen the
bare Coulomb interaction by a reduced polarization, which doesn't
contain transitions within O $p$ -- Co $d$ band complex. 

Then we construct a multi-band Hubbard Hamiltonian
\begin{equation*}
\label{eq:ham}
\operatorname{H}=\!\sum_{\mathbf{k}}
\begin{pmatrix}
&\!\!\!\mathbf{d}^{\dagger}_{\mathbf{k}}\\&\!\!\!\mathbf{p}^{\dagger}_{\mathbf{k}}
\end{pmatrix}
\!\!\begin{pmatrix}
&\!\!\!\operatorname{h}_{\mathbf{k}}^{dd}\!\!-\epsilon_{\text{dc}} &\operatorname{h}_{\mathbf{k}}^{dp}\\
&\!\!\!\operatorname{h}_{\mathbf{k}}^{pd} &\operatorname{h}_{\mathbf{k}}^{pp}
\end{pmatrix}
\!\!\begin{pmatrix}
&\!\!\!\mathbf{d}_{\mathbf{k}}\\&\!\!\!\mathbf{p}_{\mathbf{k}} 
\end{pmatrix}+ 
\!\sum_i \operatorname{W}^{dd}_i
\end{equation*}
Here $\mathbf{d}_{\mathbf{k}}$ ($\mathbf{p}_{\mathbf{k}}$) is an operator-valued vector whose elements
are Fourier transforms of $d_{i\alpha}$ ($p_{i\gamma}$), which annihilate the Co $d$ (O $p$) electron
in the orbital $\alpha$ ($\gamma$) in the $i$th unit cell. 
We have used two types of on-site interaction $\operatorname{W}^{dd}_i$:
i) the simplified one of the density-density form $w_{\alpha\beta}n_{\alpha}n_{\beta}$ and ii)
the SU(2) symmetric interaction of the general form 
$\tilde{w}_{\alpha\beta\gamma\delta}d_{\alpha}^{\dagger}d_{\beta}^{\dagger}
d_{\gamma}d_{\delta}$ in the Slater-Kanamori parametrization \cite{sk}. 
The double-counting term $\epsilon_{\text{dc}}$ approximately
corrects for the explicitly unknown mean-field part of the interaction coming from
LDA. At each DMFT iteration $\epsilon_{\text{dc}}$ is obtained  
as the the orbital average of the high-frequency self-energy, 
$\epsilon_{\text{dc}}=\overline{\Sigma(\omega\rightarrow\infty)}$,
which equals the orbitally averaged Hartree energy.
The effective Weiss field
is obtained by iterative solution of the DMFT equations \cite{dmft} using
the finite temperature Matsubara formalism and
continuous time quantum Monte-Carlo (QMC) method. The numerical demands
of solving the problem with the simplified interaction (i) \cite{ctqmc}
are substantially smaller than in the SU(2) symmetric case (ii) \cite{krylov}.
Combining the two allows us to investigate a variety of observables 
without oversimplifying the problem.
\begin{figure}
  \begin{center}
    \includegraphics[height=0.51\columnwidth,angle=270,clip]{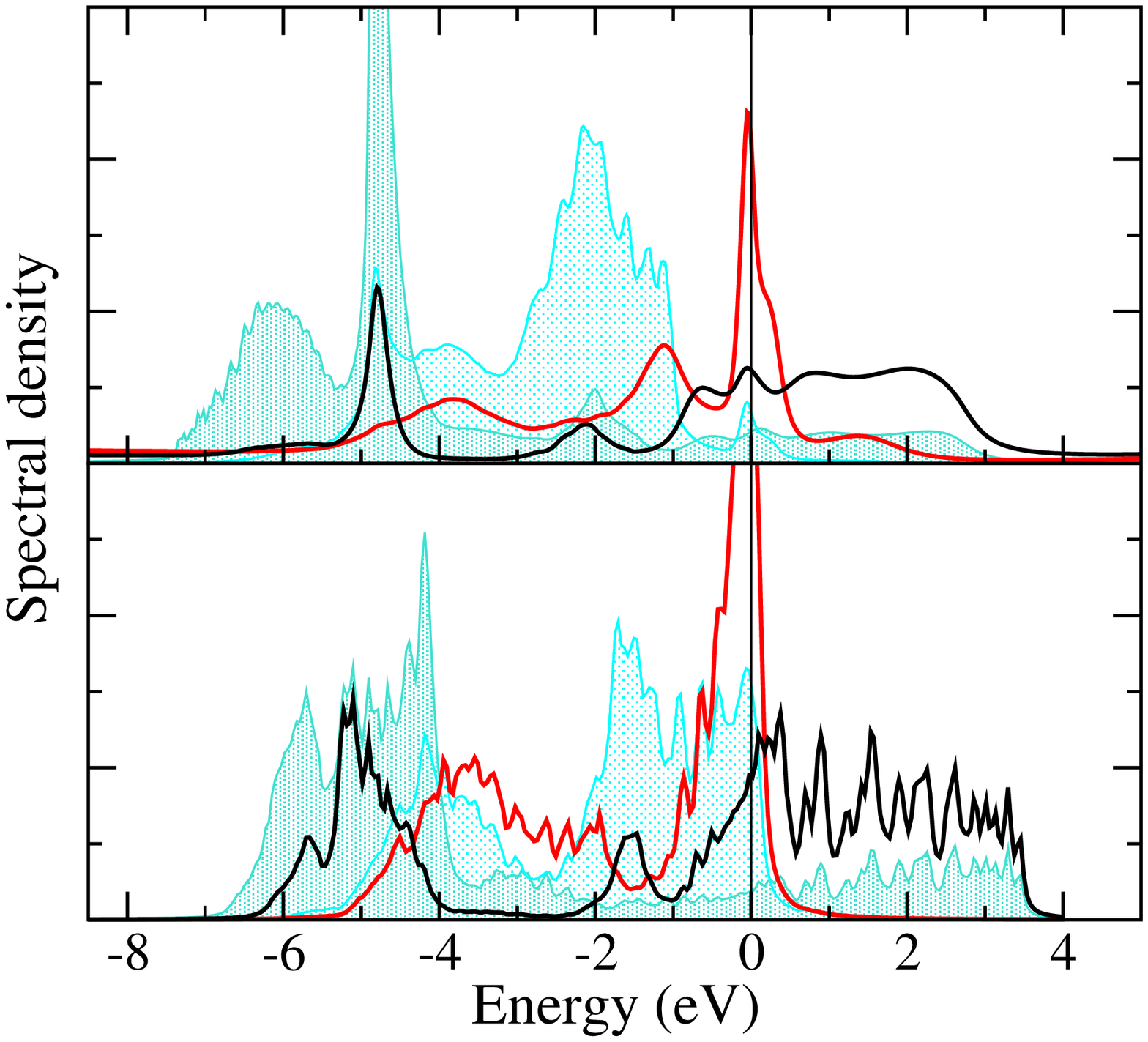}
  \includegraphics[height=0.47\columnwidth,angle=270,clip]{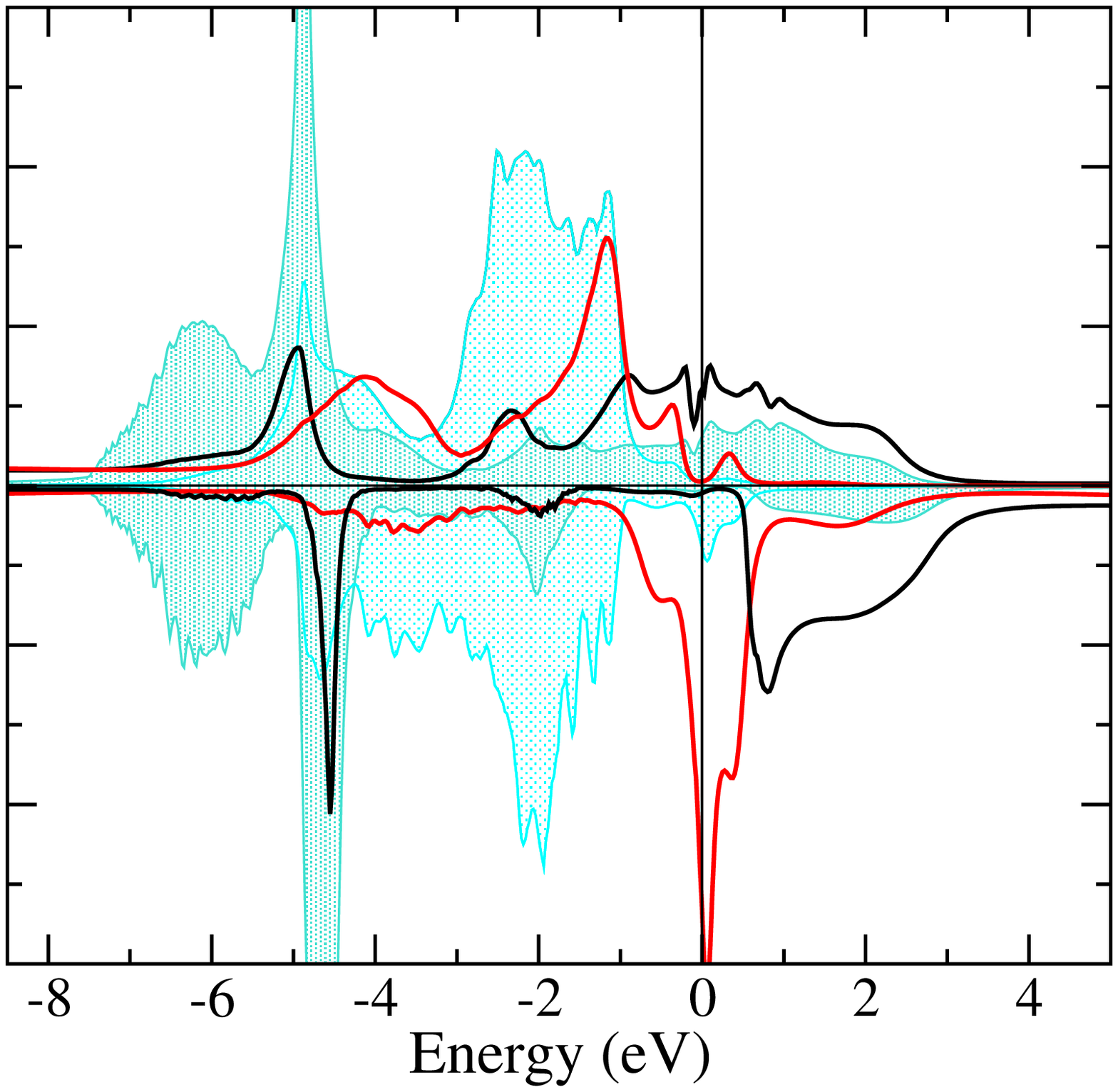}
  \end{center}
\caption{\label{fig:1} (color online) The spectral density per orbital: Co $d-e_g$ (black),
Co $d-t_{2g}$ (red), O $p_{\sigma}$ (darker blue), $p_{\pi}$ (lighter blue) in the PM DMFT (upper left),
in LDA (lower left) and in FM DMFT (right panel) solution. The minority spin in the FM phase
is distinguished by the minus sign.}
\end{figure}

First, we discuss the one-particle dynamics obtained with the density-density
interaction at $T=$1160~K. The PM solution is enforced
by the $\langle S_z\rangle=0$ constraint at each DMFT iteration.
Lifting this constraint, FM is the stable phase below 1800~K.
The desired spectral functions at real frequencies
are obtained by analytic continuation from the Matsubara
contour. Following Ref.~\onlinecite{wang09} we used the maximum entropy method \cite{maxent}
to continue the self-energy.
The Green functions were constructed from the corresponding Dyson equation. 
This approach 
provides a straightforward
access to the k-resolved spectra of both $d$ and $p$ electrons.

\begin{figure}
  \begin{center}
   \includegraphics[height=0.48\columnwidth,angle=270,clip]{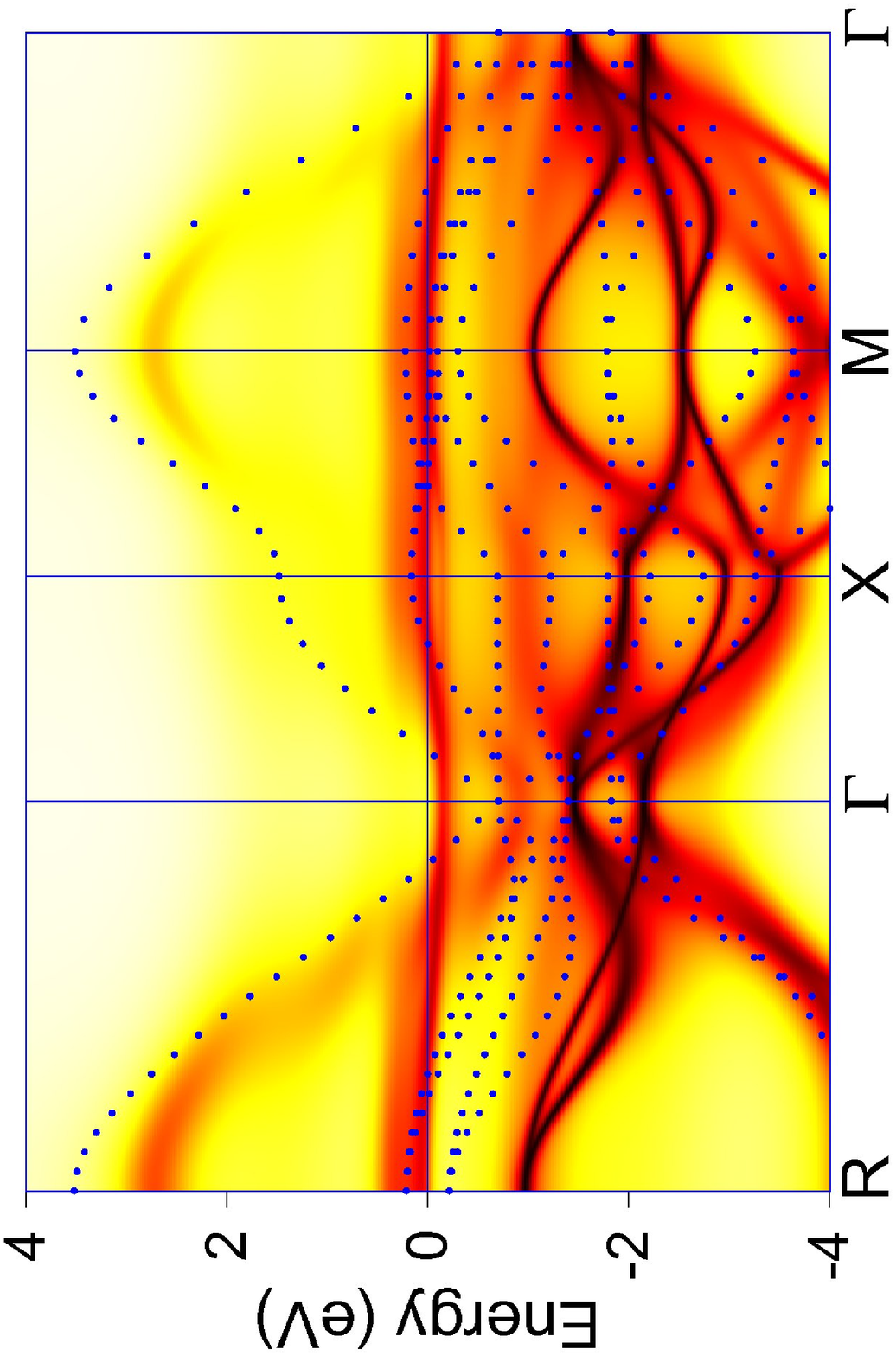}
  \includegraphics[height=0.47\columnwidth,angle=270,clip]{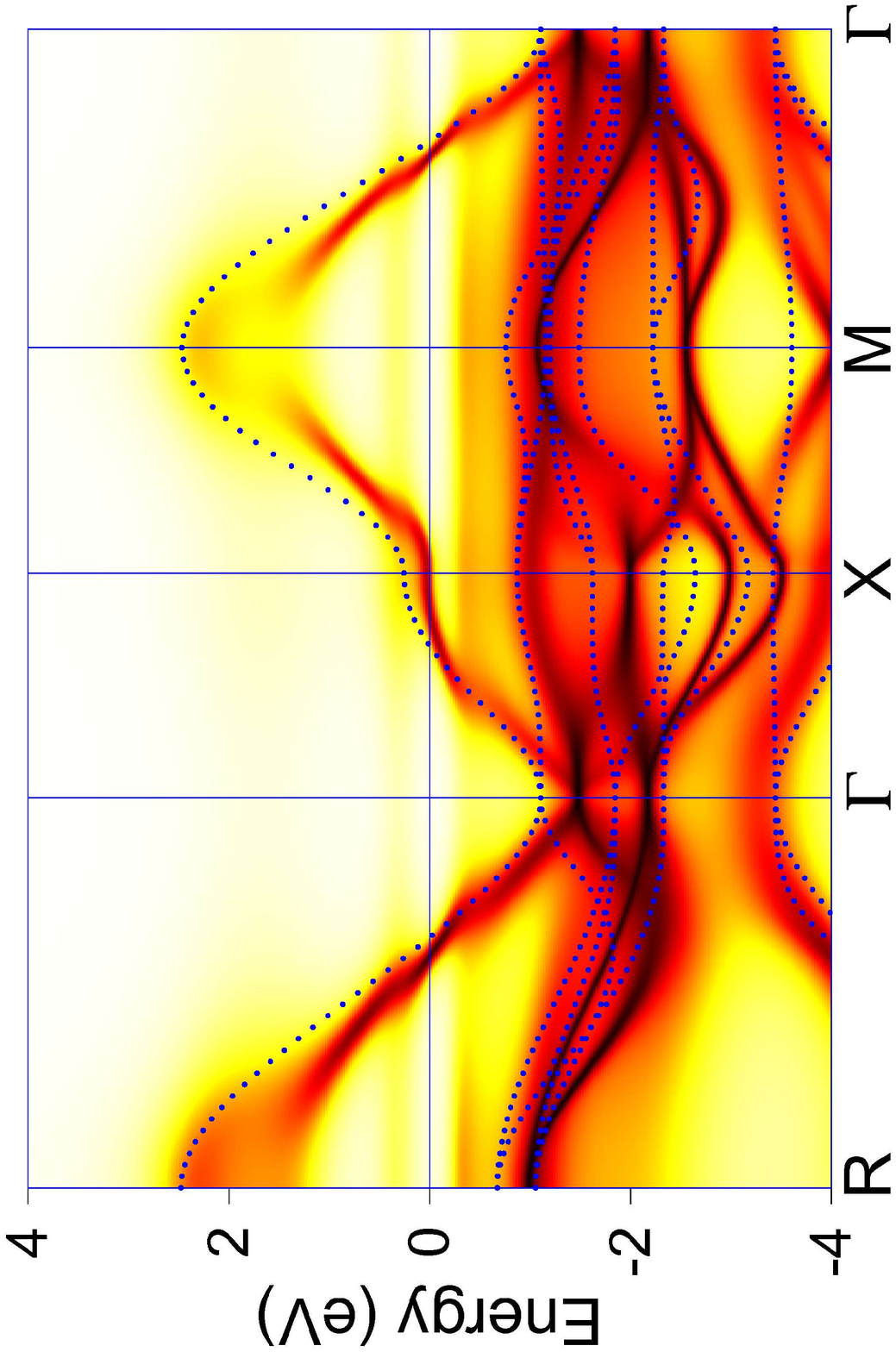}
   \includegraphics[height=0.48\columnwidth,angle=270,clip]{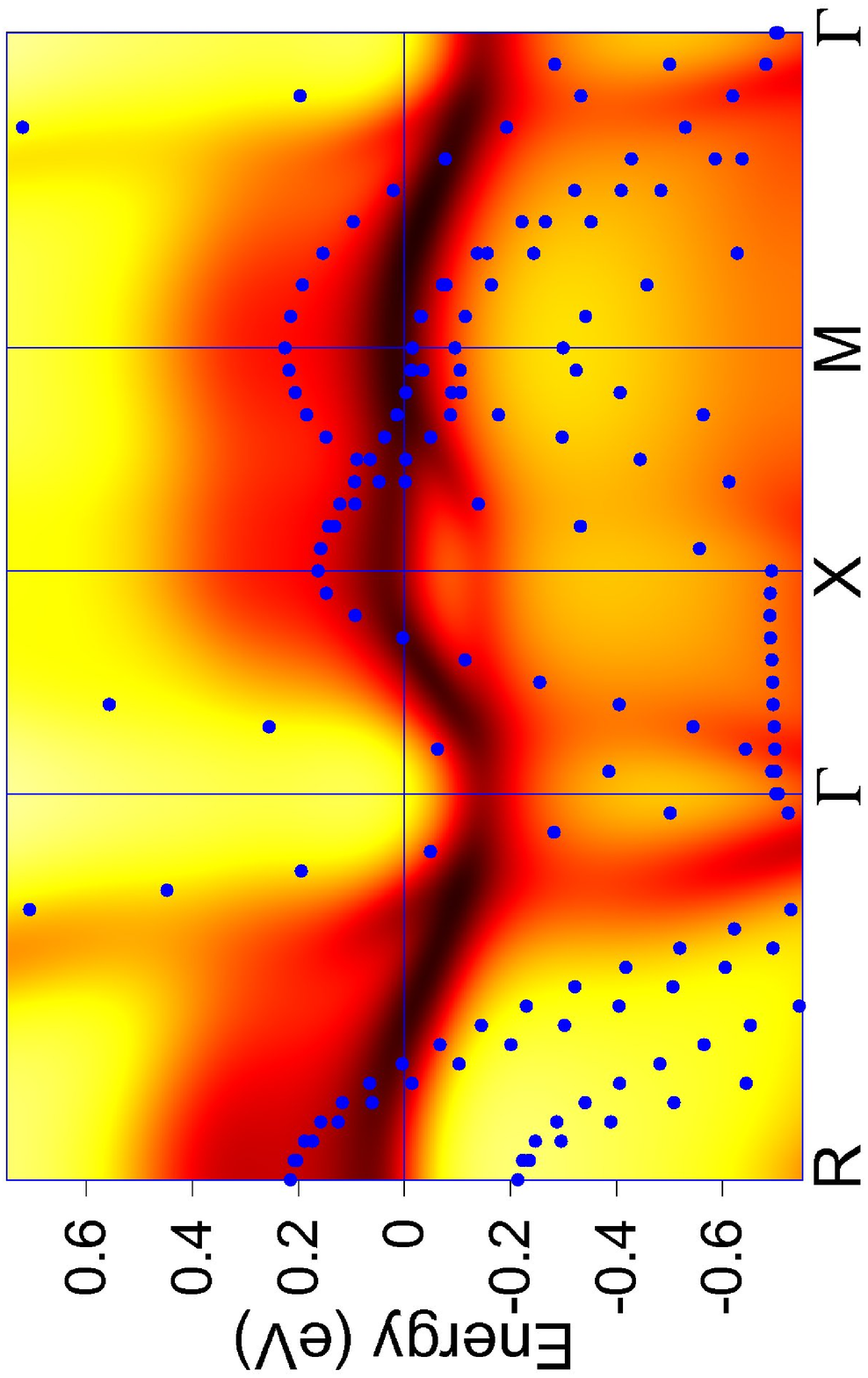}
   \includegraphics[height=0.47\columnwidth,angle=270,clip]{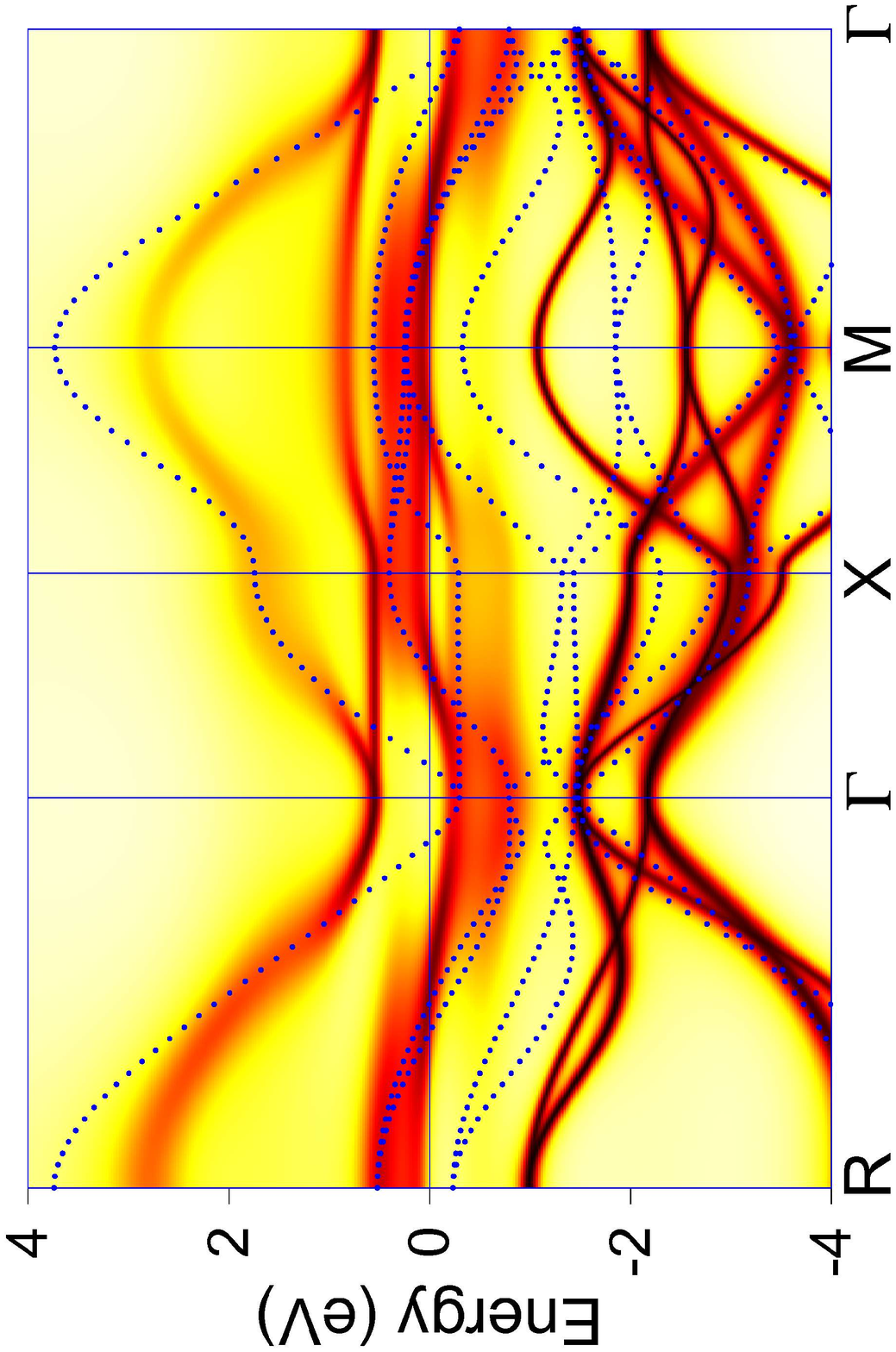}
  \end{center}
\caption{\label{fig:2} The k-resolved spectral function $A_{\mathbf{k}}(\omega)$ along the high symmetry directions
                       presented as a color plot of $A_{\mathbf{k}}(\omega)/(2+A_{\mathbf{k}}(\omega)$. The PM
                       solution is compared to the LDA bands (dots) in the upper left and in detail in the lower left panel.
                       The same spectral function in the FM phase resolved into the majority (upper right) and
                       minority (lower right) spin channels. The LSDA bands are marked with dots.}
\end{figure}
In Fig.~\ref{fig:1} we present the k-integrated DMFT spectra in the PM and FM phases
along with the LDA spectra for reference. The dynamical correlations lead to reduction of the Co $d$ occupancy from
the LDA value of 7.0 to 6.0.
Despite the strong $d-d$ interaction $\sim$10~eV the system remains metallic. Nevertheless, substantial band narrowing and
new features (including an incoherent background up to $\sim$20~eV) appear in the DMFT spectra. 
The FM phase exhibits a spin-dependent orbital polarization at the Fermi level ($E_F$) with 
the $e_g$ states dominating the majority and the $t_{2g}$ states the minority spin channel. 

\begin{figure}
   \includegraphics[height=0.65\columnwidth,angle=270,clip]{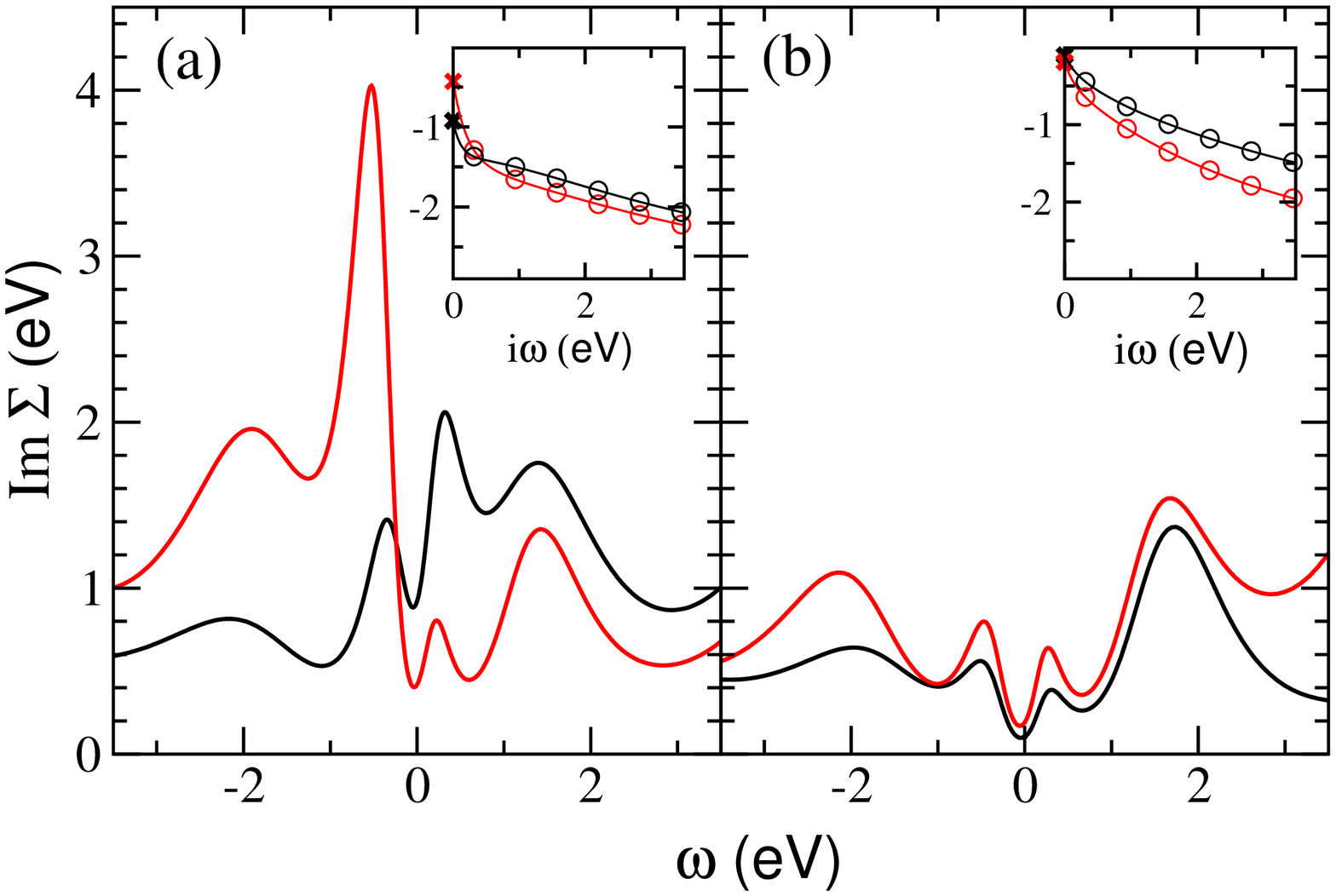}
   \includegraphics[height=0.3\columnwidth,angle=270,clip]{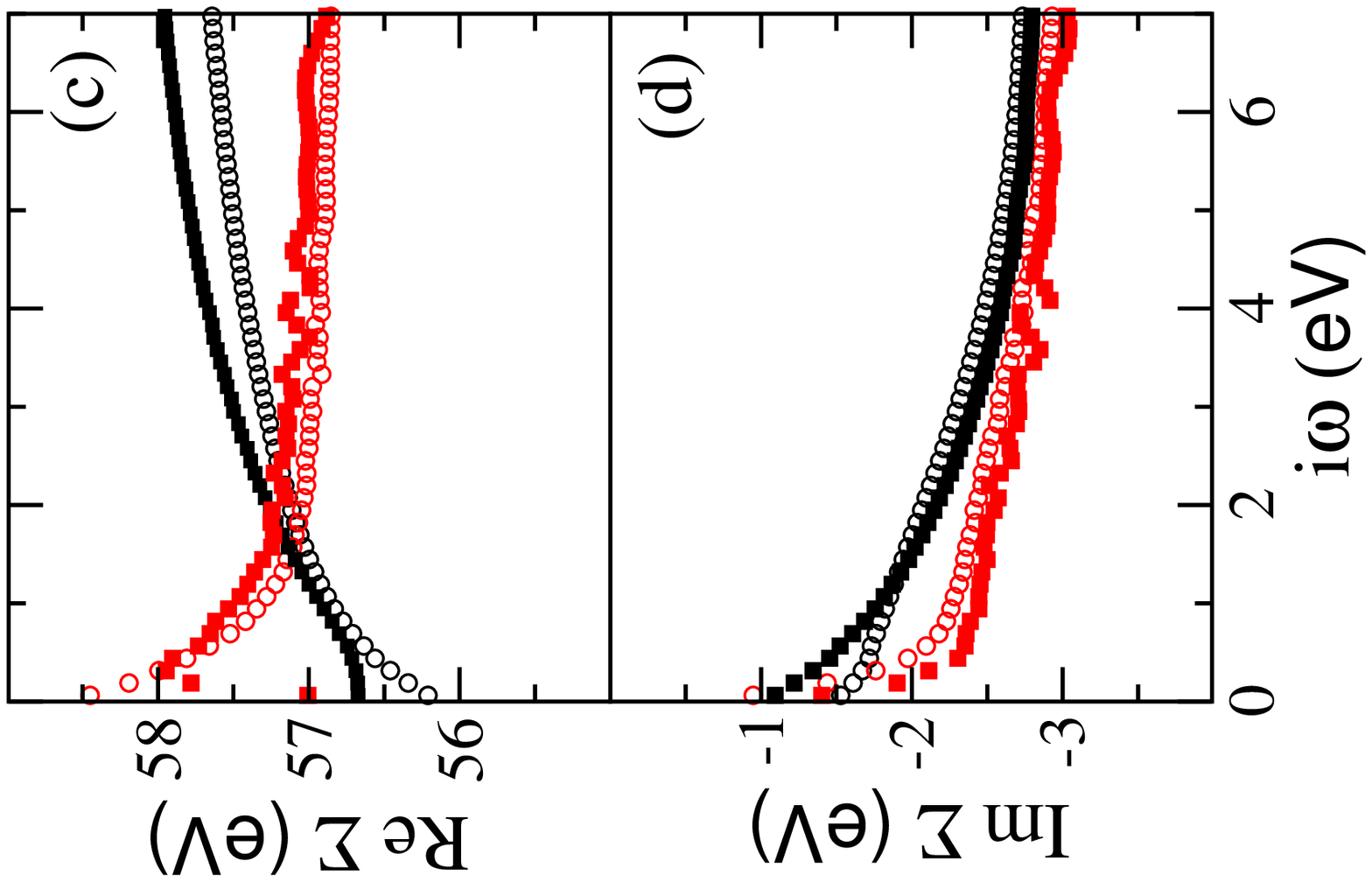}
\caption{\label{fig:3} The imaginary part of the self-energy at $T$=1160~K obtained 
       with density-density interaction in the PM (a) and FM (b) phases. 
       The diagonal $\Sigma_{ee}$ (black) and $\Sigma_{tt}$ (red) are shown (for the FM phase
       only the majority $e_g$ and minority $t_{2g}$). The insets show
       the same functions on the imaginary axes together with the values on the Matsubara 
       contour (open symbols). The zero frequency values are highlighted with the crosses.
       In panels (c) and (d) the real and imaginary parts of the self-energy obtained in the PM phase
       with the density-density (open circles) and with the SU(2) symmetric (filled squares)
       interaction at $T$=232~K are compared.
}
\end{figure}

To distinguish the band dispersion and the quasiparticle broadening
we have calculated the k-resolved spectral functions, shown in Figs.~\ref{fig:2}. 
Comparison of the PM spectra with the non-interacting LDA bands reveals
substantial differences near $E_F$. Absence of sharp quasiparticles there indicates
a strong scattering. In the FM spectra both the majority
($e_g$) and minority ($t_{2g}$) quasiparticles around $E_F$ are sharper and while
the overall band structure matches better its spin-polarized LSDA counterpart 
substantial differences around $E_F$ remain. 
The quasiparticle broadening is quantified in Fig.~\ref{fig:3}a,b
showing the imaginary part of the self-energy, which determines the quasiparticle line-width.
While both the PM and the FM self-energies exhibit a dip at $E_F$, $\operatorname{Im} \Sigma(0)$ is substantially larger
in the PM phase and the quasiparticle concept loses its meaning there. 
To assess the impact of the density-density approximation for our conclusions we have performed PM calculation
with the SU(2) symmetric interaction. In Fig.~\ref{fig:3}c,d we compare imaginary frequency self-energies at a 
temperature of 232~K, where more pronounced differences are expected.
While there is an overall agreement between the two sets of self-energies, differences exist in particular
at the lowest Matsubara frequencies. The extrapolation of the $e_g$ self-energy to zero frequency
points to a stronger scattering by local moment fluctuations in the density-density approximation. This
is plausible since Ising spins pose a bigger obstacle for quasi-particle propagation than the Heisenberg
ones as was also found in the studies on a two-band Hubbard model \cite{liebsch}. 

\begin{figure}
\includegraphics[angle=270,width=0.95\columnwidth,clip]{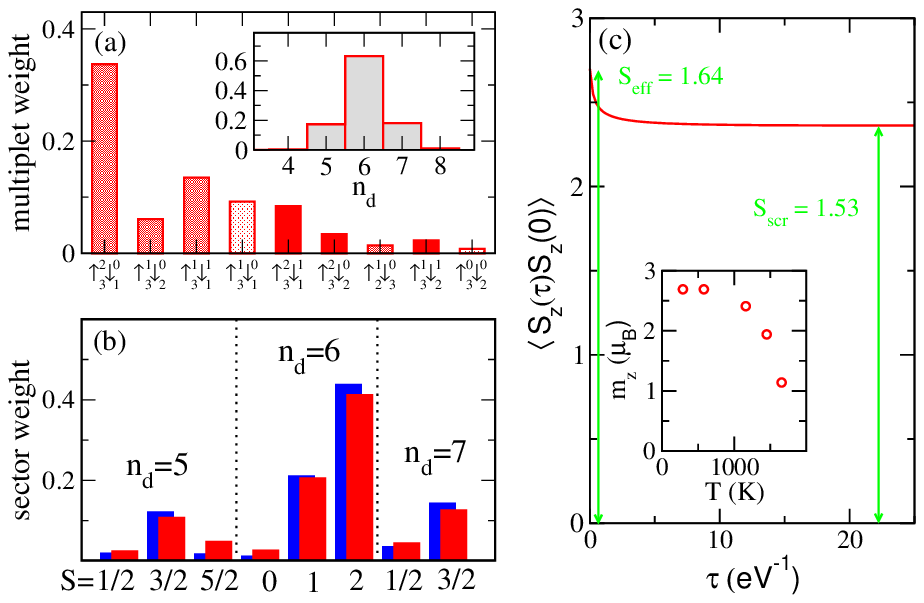}
\caption{\label{fig:mtm} 
(a) Weights of the dominant density-density multiplets 
of the Co $d$ shell at ($T$=232~K). 
The inset shows the total weights of the different charge sectors ($d^5$, $d^6$, ...),
distinguished by shading in the main panel.
(b) Comparison of the weights of different charge and spin sectors obtained
with density-density (red) and SU(2) symmetric (blue) interactions.
(c) Local spin-spin correlation in the PM phase with the density-density interaction.
The inset shows the temperature dependence of the ordered (FM) Co moment.} 
\end{figure}

The reciprocal-space quasiparticle picture is commonly used to discuss the properties of metallic systems, 
while transition metal oxides are more often described in terms of the direct-space atomic states. 
The strong coupling QMC solver is particularly useful 
in this respect 
as it provides a simple way to compute the reduced density matrix operator 
for the interacting atom (called state statistics in Ref.~\cite{werner07}). This quantity, which 
tells `how much time the atom spends in a particular many-body state', allows straightforward evaluation
of the expectation value of any local operator. In the inset of Fig.~\ref{fig:mtm}a we show the 
the statistics of the valence states ($d^5$, $d^6$, ...) as well as 
the atomic multiplets with the largest weights. Contrary to the formal Co$^{4+}$ ($d^5$) valence, 
the Co atoms spend most of the time in $d^6$ configuration with almost 
symmetric fluctuations to $d^5$ and $d^7$, a feature 
consistent with metallic behavior \cite{ylvi,mcma}. 
Inspecting the weights of different multiplets (Fig.~\ref{fig:mtm}a) we find the $d^6$ HS state ($\uparrow^2_3\downarrow^0_1$ where
the subscripts stand for the number of $t_{2g}$ and superscripts for the number of $e_g$ electrons)
to be most abundant with about 33~\%. Since 
the density-density and SU(2) symmetric interaction leads to different multiplet structures 
we cannot compare directly the multiplet weights. Nevertheless, we can compare total
weights of sectors indexed by charge and the total spin $|S_z|$ (density-density) or $S=\sqrt{S_x^2+S_y^2+S_z^2}$
(SU(2)), shown in Fig.~\ref{fig:mtm}b,
and find a close match.
We attribute it to the fact that the dominant HS state is well captured with both forms
of the interaction. This does not necessarily mean that all physical quantities are similar.
Differences are expected in particular for quantities to which the individual members of the multiplets
contribute differently such as the spin susceptibility and thus the FM $T_C$, which are certainly overestimated 
with the density-density
interaction. \cite{fe2o3,dd_comp}
To resolve the question of the origin of the local moment
we turn our attention to the local spin susceptibility.
Large weight of the HS state by itself does not guarantee a Curie-Weiss susceptibility, $\chi\sim1/T$.
It is necessary that the moments are long lived as is clear from the Kubo formula
$\chi=\int_0^{\beta=1/T}d\tau \langle S_z(\tau)S_z(0)\rangle$.
The spin-spin correlation function along the imaginary time contour obtained with
the density-density interaction is
shown in Fig.~\ref{fig:mtm}. The flatting out after the initial rapid decay is indicative of the local moment behavior.
The PM Co moment of 3.06~$\mu_B$, estimated by $S_{\text{scr}}=\langle S_z(\beta/2)S_z(0)\rangle$, is close to 
the saturated FM moment on the Co atom of 2.7~$\mu_B$ (which is fairly close
to the LSDA value of 2.58~$\mu_B$ per u.c. and the experimental saturation magnetization 
of 2.5~$\mu_B$ \cite{long2011}), shown in the inset. 
To make contact with
the SU(2) symmetric interaction
we compare the effective moments derived from the instantaneous correlators $\langle S_z^2\rangle_{DD}=S_{\text{eff}}^2$ and
$\langle S^2\rangle_{SU(2)}=S_{\text{eff}}(S_{\text{eff}}+1)$.
$S_{\text{eff}}$ of about 1.64 and 1.61 are obtained in the density-density and SU(2) symmetric
calculation, respectively.

The experimental PM moments are often associated with a particular
multiplet and a fractional value is interpreted as a mixture of contributions from different
multiplets, the thermally induced PM susceptibility in LaCoO$_3$ being
a textbook example. 
In general, however, it is not possible to divide the susceptibility into multiplet contributions
as can be shown by expressing 
the spin operators $S_z$ in terms of the atomic states projectors $P_A$  
\begin{equation*}
\chi=\sum_{A,B}S_z(A)S_z(B)\int_0^{1/T}\!\!\!\!\!\!d\tau\langle P_A(\tau)P_B(0)\rangle,
\end{equation*}
with $A$ and $B$ running over all the atomic many-body states. Contributions
of individual multiplets make sense only if the
matrix $\Pi_{AB}=\int_0^{\beta}d\tau\langle P_A(\tau)P_B(0)\rangle$ has negligible
off-diagonal elements. This means that causal evolution
between such states has low probability and their simultaneous population is a result
of ensemble averaging. A trivial example is an isolated atom for which $\Pi_{AB}$ 
is diagonal in the basis of atomic eigenstates. It is also possible that $\Pi_{AB}$ has a block structure.
In that case each block can be associated with the dominant multiplet
and fluctuations around it, arising for example from hybridization with ligands. 
Examples of such behavior can be found in systems
near the high-spin--low-spin transition \cite{werner07,kunes-krapek}
or systems with fluctuating valence \cite{ylvi}. 

\begin{figure}
   \includegraphics[width=0.95\columnwidth,clip]{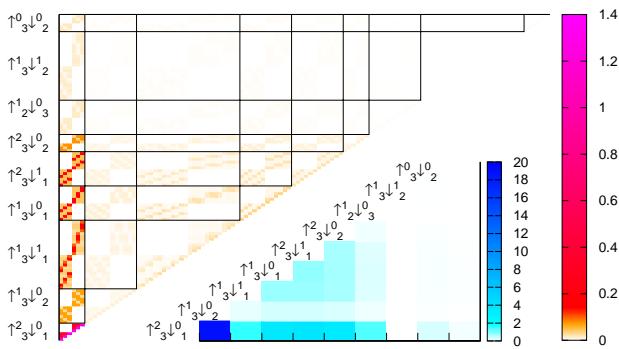}
\caption{\label{fig:states_corr} The atomic state correlation matrix $\Pi_{AB}$ between 
the multiplets shown in Fig.~\ref{fig:mtm} obtained with density-density interaction at
232~K.
Upper left triangle shows the state-by-state relative contributions.
The lower right triangle shows the contribution 
of multiplet pairs to the local susceptibility.} 
\end{figure}

The state correlation matrix $\Pi_{AB}$ obtained with density-density interaction (SU(2) symmetric calculation
is prohibitively expensive)
 is presented in Fig.~\ref{fig:states_corr} with
only the multiplets with large weights shown for simplicity. The depicted part of $\Pi_{AB}$ 
contains 70\% of the total $\sum_{A,B} \Pi_{AB}=\beta$.
Only states with the same spin orientation exhibit sizable off-diagonal elements, leading to an apparent block structure 
within each black rectangle. Other than this no block structure can be found which means
that all the states are connected by causal evolution. 
Therefore we cannot view the system as a mixture of HS and IS states and distinguish their
contributions to the susceptibility. This statement is quantified in the lower right
panel of Fig.~\ref{fig:states_corr} where the contributions of the different multiplet pairs to the 
spin susceptibility are shown. Although the HS-HS contribution is largest it amounts
only to 23~\% of the total.

There is a connection between the multiplet picture and the one-particle spectra.
Even in the PM phase there exists a local spin orientation which is long
lived, i.e.~there is a strong correlation between the states of this orientation.
The local-majority $t_{2g}$ orbitals remain filled, which becomes
apparent in the FM spectra. Similarly the minority $e_g$ states are mostly empty
except for short visits of the $\uparrow^1_3\downarrow^1_1$ configuration, the short life-time of which  
can be deduced from the relatively small diagonal element of $\Pi_{AB}$. These fluctuations
are responsible for a saturated Co-O $\sigma$-bond and do not lead to metalicity.
Based on these observations it is plausible to ascribe the FM order to the double exchange mechanism. 
Unlike in manganites we cannot describe the systems as local $t_{2g}$ moments plus 
itinerant $e_g$ electrons. In SrCoO$_3$ the inter-site exchange is mediated by both the majority $e_g$
and minority $t_{2g}$ electrons. The minority $e_g$ electrons participate in 
Co-O bonding, but are not active in the double exchange.
The calculated transition temperature substantially overestimates the experimental one, while
the size of the ordered moment agrees rather well.
There are two obvious deficiencies of our theory, the Ising character
of the local moments and the lack of inter-site spin-spin correlations. To assess their
importance a more complete theory would be necessary. 

Previous theoretical studies \cite{potze95,zhuang98} concluded that IS state dominates the ground state
of SrCoO$_3$. We pass over the fact that these studies used phenomenological parameters 
while we are using full `first-principles' bandstructure and focus on the qualitative aspects.
Zhuang {\it al.} \cite{zhuang98} used the unrestricted Hartree-Fock method, which allows the 
system to settle in a particular atomic state, but does not allow quantum or thermal
fluctuations and thus cannot properly describe the competition between electron localization and
itinerancy. DMFT does not have these deficiencies and can be considered a systematic
improvement over the Hartree-Fock approximation. In the other study, Potze {\it et al.} \cite{potze95} used
exact diagonalization on a small cluster.
They found an IS {\it cluster} ground state with the dominant (67~\%) contribution
formed by locking the $d^6$ {\it atomic} HS state on Co, similar to our results,
with an anti-ferromagnetically oriented ligand hole.
Formation of a bound $d^6\underline{L}$ state with conserved total spin is inevitable 
in a cluster, because the ligand hole has only one Co partner from which it cannot run away
and thus is strongly correlated with its state.
The situation in a metallic system may be quite different and the Co-O correlation 
obtained from the cluster calculation is certainly exaggerated. On the other hand,
in DMFT the dynamical Co-O correlation is completely neglected, i.e. the Co atom senses
the same average environment with $\sim1/3$ of a hole per O atom irrespective of its own instantaneous state.
In the FM phase some Co-O correlation becomes static, which is reflected in 
the O $p_{\sigma}$ orbitals being polarized opposite to the net magnetization ($-$0.03$\mu_B$ at 1160~K).
The total O polarization is positive due to the $p_{\pi}$ orbitals (0.04$\mu_B$ per orbital).

In conclusion, using LDA+DMFT approach we have found that SrCoO$_3$ is a FM metal with 
transition temperature in the hundreds K range. In the FM phase the majority (minority) spin states at
the Fermi level exhibit a complete $e_g$ ($t_{2g}$) polarization.  
Reduced quasi-particle scattering in the FM phase with respect to the PM one points
to a negative magneto-resistance effect, but actual transport calculations were not performed.
The statistical operator projected on the Co atom is dominated by the $d^6$ HS state, which is reflected in local
fast-probe experiments such as the x-ray absorption \cite{potze95}.
On the other hand, the long-time properties such as the static susceptibility are affected
by the causal evolution connecting the different atomic multiplets and thus
cannot be inferred from the multiplet weights.
Taking into account the proper spin-rotation symmetry of the local interaction alters 
some quantitative details and clearly confirms the overall picture.

We thank Z. Jir\'ak, P. Nov\'ak and K. Held 
for numerous discussions. N. P. and G. S. are indebted to M. Ferrero, E. Gull and P. Werner for help and 
feedback in writing the SU(2)-symmetric code. This work was supported by
the Grant No. P204/10/0284 of the Grant Agency of the Czech Republic
and by the Deutsche Forschungsgemeinschaft through FOR 1346 (project ID I597-N16 of the Austrian Science Fund, AT).

\end{document}